# PERANGKAT LUNAK BANTU MENGENAL HURUF ARAB MELAYU KE BENTUK HURUF LATIN BAHASA INDONESIA

**Oleh: Nuril Aini, Leon Andretti Abdillah, & Jemakmun**
**Mahasiswa & Dosen Universitas Bina Darma, Palembang**

***Abstracts**: The development of computer science has contributed greatly for increasing of efficiency and effectivity. Many areas are covered by computer science, included education. The purpose of this research is to introduce jawi a type of Indonesian letters. Jawi's letter is one of the most popular letter in the past. But right now few people can read and understand it. Many documents in the past was written in Jawi. Tthe writer develop/build the software using Pressman method, and tools such as Microsoft Visual Basic, and Microsoft Access. This software can introduce Jawi then people can learn it easily.*

## 1. PENDAHULUAN

Dalam era teknologi yang semakin maju sekarang ini, pemanfaatan komputer semakin dirasakan untuk dapat mempermudah pekerjaan, karena setiap informasi yang dihasilkan akan selalu ditampilkan kedalam bentuk informasi yang akurat dan spesifik, baik itu pekerjaan dalam bidang pendidikan, ekonomi, manajemen, sejarah mapun terapan ilmu lainnya. Hal ini dikarenakan komputer lebih unggul dari pada alat hitung lain karena mampu mengingat data, mengambil keputusan yang logis, mengelola data dan informasi. Kini komputer lebih tepat dinamakan pengelola data elektronik.

Huruf arab melayu merupakan salah satu bentuk penulisan yang sering digunakan untuk menyampaikan sesuatu berita kepada orang lain yang ditulis dalam bentuk huruf arab tanpa menggunakan eja-an. Penulisan huruf arab melayu ini sedikit berbeda dengan penulisan huruf arab yang terdapat dalam kitab suci umat beragama islam, yaitu Al-Qur'an. Huruf arab melayu bukan merupakan bentuk bahasa yang resmi digunakan oleh rakyat Indonesia, dikarenakan penggunaan huruf arab melayu ini semakin berkurang dan untuk membacanya dibutuhkan seseorang yang juga mampu membaca huruf arab melayu ini, juga dikarenakan penulisan bentuk huruf arab melayu ini tidak dimasukkan kedalam kurikulum pendidikan di Indonesia. Namun banyak digunakan untuk penulisan kitab-kitab.

Untuk membaca huruf arab melayu ini, sebaiknya mempelajari lebih dahulu huruf arab yang terdapat dalam kitab suci Al-Qur'an, hal ini dikarenakan bentuk penulisanya sangat mirip dengan penulisan huruf arab yang terdapat dalam kitab suci Al-Qur'an walaupun terdapat sedikit perbedaan dalam bentuk penulisannya. Karena dengan mempelajari kitab suci Al-Qur'an terlebih dahulu, maka kesulitan dalam hal memperlajari huruf arab melayu ini sedikit dapat membantu.

Berbagai kesulitan yang didapat untuk menulis maupun cara membaca huruf arab melayu ini yaitu tulisan arab melayu ini tidak menggunakan ejaan atau harkat dan huruf – huruf yang di pakai dalam penulisan arab melayu ini ada bebrapa huruf yang berbeda dari huruf hijaiyah sebagai huruf tambahan dan selain itu juga tempat untuk mempelajarinya sekarang ini sulit didapat oleh karena itu bila ingin mempelajari cara membaca dan menulis arab malayu ini terlebih dahulu harus mempelajari huruf – huruf yang ada di Al qur'an dan cara membacanya. Salah satu contoh bentuk penulisan bentuk huruf arab melayu ini terdapat dalam buku yang berjudul khasiat surat Alfatihah dan ayat Al-Kursi.

Kesulitan yang dirasakan tersebut menghasilkan sebuah ide bagi penulis bagaimana caranya membantu mempermudah cara membaca huruf arab malayu ini yang dituangkan kedalam suatu perangkat lunak Bantu, yaitu mengganti cara konensional tersebut menjadi cara yang lebih efektif dalam mengenal dan membaca huruf arab melayu menggunakan program aplikasi *Microsoft Visual Basic* versi 6.0.

Berdasarkan uraian yang telah dijelaskan diatas dan informasi yang diperoleh penulis mengenai cara penulisan dan membaca huruf arab melayu ini, maka masalah yang dibahas dapat didefinisikan sebagai berikut: "Bagaimana membangun perangkat lunak bantu mengenal huruf arab melayu ke bentuk huruf latin bahasa Indonesia menggunakan program aplikasi *Microsoft Visual Basic* versi 6.0."

Batasan Masalah, Dalam penyusunan proposal ini penulis membatasi permasalahan pada mengenal bentuk huruf arab melayu ini kedalam bentuk huruf latin bahasa Indonesia berdasarkan bentuk kata dan kalimat yang dimasukkan.

Tujuan Penelitian, merancang dan membangun perangkat lunak Bantu mengenal cara penulisan dan membaca huruf arab melayu ini dengan menggunakan program aplikasi *Microsoft Visual Basic* versi 6.0.

Manfaat Penelitian: 1) Perangkat lunak ini diharapkan dapat dijadikan sebagai salah satu alternatif untuk mengenal huruf arab melayu ini, dan 2)

Diharapkan lebih mempermudah mempelajari huruf arab melayu, sehingga dapat mengerti bagaimana cara penulisan dan pembacaan huruf arab melayu.

## 2. TINJAUAN PUSTAKA

### Perangkat Lunak Bantu

Menurut Pressman (2002:10) perangkat lunak adalah: 1) intruksi (Program komputer) yang bila di *eksekusi* dapat menjalankan fungsi tertentu, 2) struktur data yang dapat membuat program memanipulasi informasi, 3) dokumen yang menjelaskan operasi dan penggunaan program. Sedangkan Abdillah (2006:5) mengemukanan Perangkat Lunak komputer merupakan perangkat yang secara nyata tidak dapat di-akses oleh panca indera manusia, namun ia ada dan sangat penting peranannya. Jadi perangkat lunak merupakan program-program komputer (tidak dapat di-akses langsung oleh panca indera manusia) yang berguna untuk menjalankan suatu pekerjaan sesuai dengan yang diinginkan. Program tersebut ditulis dengan bahasa khusus yang dimengerti oleh komputer.

### Huruf Arab Melayu (Jawi)

Jawi adalah salah satu bentuk tulisan kuno yang digunakan oleh rakyat melayu, khususnya yang ditulis dengan menggunakan tulisan huruf arab melayu. Rakyat melayu menggunakan bahasa ini untuk bekerja sama, berinteraksi, dan mengidentifikasikan diri (Tim Penyusun, 1997:77).

Huruf **Jawi** adalah sebuah sistem tulis yang sudah berabad-abad lamanya di Nusantara. Kemunculannya berkait secara langsung dengan kedatangan agama Islam ke Nusantara. Tulisan Jawi berasal dari tulisan Arab dan merupakan huruf Arab yang dimasukkan dalam sistem penulisan bahasa Melayu. Bukti terawal tulisan Jawi ini berada di Malaysia adalah dengan adanya Prasasti Terengganu yang bertarikh 70 Hijriah atau 13 Masehi. Tarikh ini agak problematis sebab bilangan tahun ini ditulis, tidak dengan angka. Di sini hanya bisa terbaca *tujuh ratus dua*: 702H. Tetapi kata *dua* ini bisa diikuti dengan kata lain; (20-29) atau -*lapan* -> *dualapan* -> "delapan". Kata ini bisa pula diikuti dengan kata "sembilan". Dengan ini kemungkinan tarikh ini menjadi banyak: (702, 720 - 729, atau 780 - 789 H). Tetapi karena prasasti ini juga menyebut bahwa tahun ini adalah "Tahun Kepiting" maka hanya ada dua kemungkinan yang tersisa: tahun 1326M / 1386M.

| | | | | | | |
|---|---|---|---|---|---|---|
| خ | ح | ج | ث | ت | ب | ا |
| ش | س | ر | ز | ر | ذ | د |
| ڤ | غ | ع | ظ | ط | ض | ص |
| ه | و | ن | م | ل | ك | ق |
| | | ڽ | ڠ | چ | ݢ | ى |

**Gambar 1. Huruf Jawi Dasar**

Adapun contoh penggunaan huruf arab melayu ke bentuk huruf latin bahasa Indonesia, sebagai berikut: Huruf Arab Melayu → (ڤوروڤ ابرا ويلم). Latin Bahasa Indonesia → (ايسينودنا نيتلا سحاب). (http://ms.wikipedia.orgwiki/ Tulisan_Jawi)

**Tabel 1. Contoh Pembacaan Huruf Jawi Dasar**

| Huruf Jawi | Contoh Pembacaan | |
|---|---|---|
| ا | Api | اڤي |
| ب | Batu | باتو |
| ت | Titi | تيتي |
| ث | Selasa | ثلاس |
| ج | Jari | جاري |
| ح | Khusus | خصوص |
| خ | Hidup | حيدوڤ |
| د | Dadu | دادو |
| ذ | Zat | ذات |
| ر | Ratu | راتو |
| ز | Zirapah | زيراڤه |
| س | Satu | ساتو |
| ش | Syling | شيليڠ |
| ص | Sabar | صابر |
| ض | Wudhu | وضوق |
| ط | Bathin | بطين |
| ظ | Zahir | ظهير |
| ع | Ilmu | علمو |
| غ | Ghoib | غيب |
| ڤ | Pikir | ڤيكير |
| ق | Qori | قولط |
| ك | Kamu | كمو |
| ل | Lisan | ليسان |
| م | Makan | مكن |
| ن | Nuri | نوري |
| و | Wayang | وايڠ |
| ه | Hati | حاتي |
| ى | Yakin | يكين |
| چ | Cari | چري |
| ڠ | Nganga | ڠاڠ |
| ݢ | Gigi | ݢيݢي |
| ڽ | Niru | ڽيرو |

**Teknik Pembacaan Huruf Arab Melayu**

Sistem penulisan JAWI berbeda dengan sistem penulisan RUMI kerana ia bermula dari kanan ke kiri. Perhatikan gambar rajah di bawah.

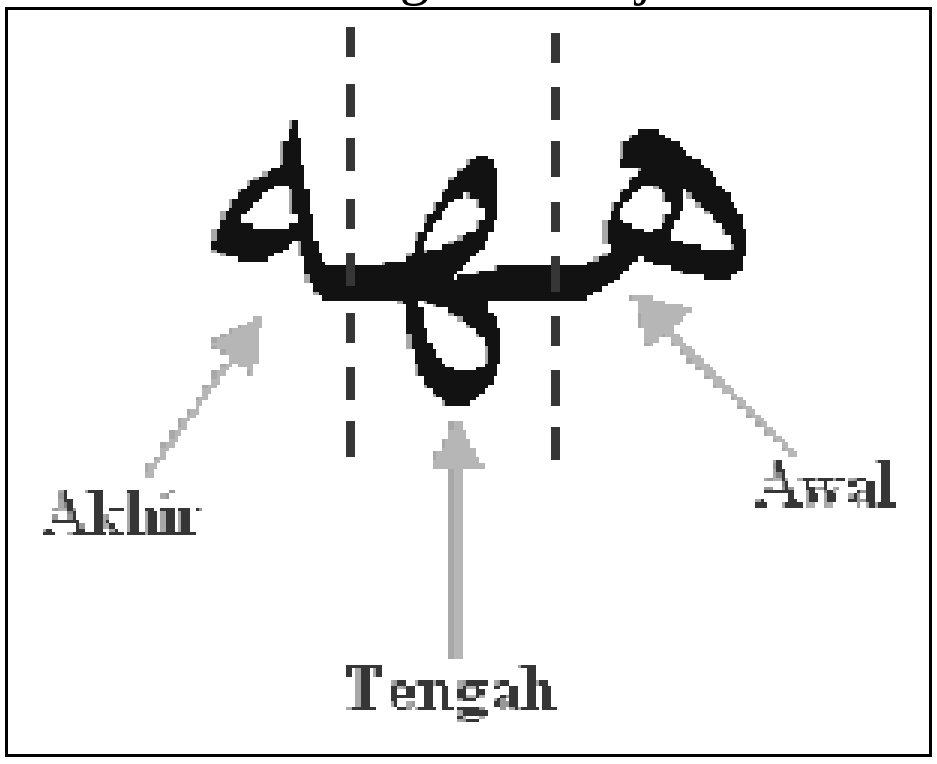

**Gambar 2. Teknik Penulisan Huruf Jawi**

Jika sesuatu huruf tidak boleh disambungkan dengan huruf lain selepasnya, ia perlu ditulis mengikut cara penulisan huruf di awal.

**Huruf Vokal**

Terdapat 3 (tiga) huruf vokal di dalam sistem penulisan Jawi, yaitu: 1) huruf Alif (ا), Waw (و) dan Ya (ي). (http://id.wikipedia.org/wiki/Huruf_Jawi)

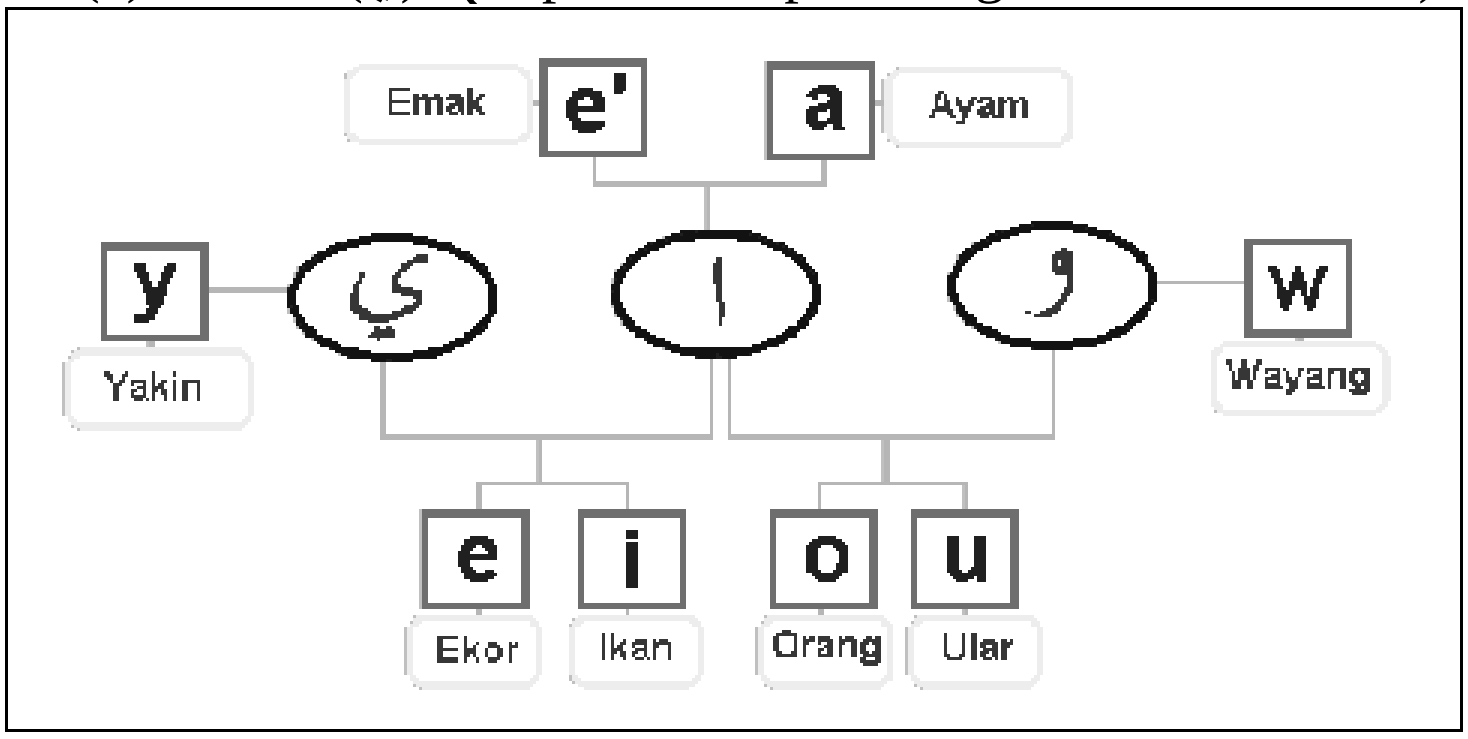

**Gambar 3. Rajah Huruf Jawi Vokal**

Huruf Alif dan Waw boleh bergabung menghasilkan bunyi 'O' untuk 'Orang' dan 'U' untuk 'Ular' Contoh: a) Bunyi 'O' > Orang = (wاورا), dan b) Bunyi 'U' > Ular = (اولر).

Huruf Alif dan Ya boleh bergabung menghasilkan bunyi 'E' untuk 'Ekor' dan 'I' untuk 'Ikan' (Smiliar, 2004). Contoh: a) Huruf Alif mempunyai dua bunyi yaitu 'A' untuk 'Ayam' dan 'è' untuk 'Emak', b) Huruf Waw mempunyai bunyi 'Wa' untuk 'Wayang', dan c) Huruf Ya mempunyai bunyi 'Ya' untuk 'Yakin' .

Huruf Alif dan Waw boleh bergabung menghasilkan bunyi 'O' untuk 'Orang' dan 'U' untuk 'Ular'. Huruf Alif dan Ya boleh bergabung menghasilkan bunyi 'E' untuk 'Ekor' dan 'I' untuk 'Ikan'. Contoh: a) Bunyi 'E' > Ekor = (ايكور), dan b) Bunyi 'I’ > Ikan = (ايكن) Perhatikan ringkasan rajah huruf vokal pada gambar 3.

**Huruf Konsonan**

Selain ketiga huruf Alif, Waw dan Ya ada huruf konsonan yaitu:

| |
|---|
| ب ت ث ج ح خ د ذ ر ز س ش ص ض ط ظ ع غ ف ق ك ك ل م ن ه ع |

**Gambar 4. Rajah Huruf Jawi Konsonan**

**Huruf Latin**

Abjad Latin adalah huruf yang pertama kalinya dipakai oleh orang Romawi menuliskan bahasa latin kira-kira abad ke 7 sebelum Masehi. Mereka belajar menulis dari orang-orang Etruska. Sedangkan orang Etruska belajar dari orang Yunani. Aksara Etruska merupakan adapatasi dari abjad Yunani.

Pada saat ini abjad Latin adalah huruf yang paling banyak dipakai di dunia untuk menuliskan berbagai bahasa, termasuk bahasa Indonesia. Orang Romawi hanya membutuhkan 23 huruf untuk menulis bahasa latin:

**A B C D E F G H I K L M N O P Q R S T V X Y Z**

Orang Romawi hanya menggunakan huruf besar saja. Huruf kecil baru berkembang kemudian hari, dari bentuk kursif tulisan tangan. Huruf I dan V, bisa dipakai sebagai vokal dan konsonan. Sedangkan huruf K, X, Y, dan Z hanya digunakan untuk menulis kata-kata pungutan dari bahasa Yunan.

Huruf C kemungkinan dilafazkan seperti dalam bahasa Indonesia. Lafaz di kemudian hari di mana C di depan i dan e dilafazkan sebagai [s] dan di depan vokal lainnya sebagai [k], kemungkinan besar tidak berlaku zaman dahulu. Tetapi ada pula yang berpendapat bahwa C di semua kasus diucapkan sebagai [k]. Huruf-huruf **J**, **U** dan **W** ditambahkan di kemudian hari untuk menuliskan bahasa non-Latin lainnya, terutama bahasa-bahasa Jermanik. J adalah sebuah varian I, dan dilafazkan seperti /y/ dalam bahasa Indonesia. Sedangkan huruf U adalah varian V, dan W diperkenalkan sebagai "v ganda" untuk membedakan bunyi /v/ dan /w/ yang tidak relevan dalam bahasa Latin. (http://id.wikipedia.org/wiki/Huruf_Jawi)

### Teknologi Informasi dalam Pendidikan

Penggunaan komputer dalam pendidikan akan semakin meluas bila harga komputer teerjangkau oeh masyarakat luas, lebih canggih dan serba guna penggunaan komputer di sekolah, universitas, lembaga kursus, rumah dan diberbagai lembaga pendidikan menunjukkan masyarakat kita memiliki elemen komputer yang dapat memenuhi kebutuhan pendidikan dan sebagai fasilitas dalam belajar (Hartono, 1992:14).

### Microsoft Visual BASIC

*Microsoft Visual Basic* adalah bahasa pemrograman yang digunakan untuk membuat aplikasi *windows* yang berbasis grafik (GUI – *Graphical User Interface*). *Visual Basic* merupakan *event-driven programming* (pemrograman terkendali kejadian) artinya program menunggu sampai adanya respon dari pemakai berupa *event* (kejadian) tertentu, seperti tombol ditekan, menu pilihan dan lain-lain. Ketika *event* terdeteksi, kode yang berhubungan dengan *event* (*procedure event*) akan dijalankan. (Suryo, 2000:1).

## 3. METODOLOGI PENELITIAN

### Lokasi Penelitian

Penelitian dilakukan di Perpustakaan Universitas Bina Darma Palembang yang beralamat di Jl. A. Yani Palembang dan waktu penelitian dilaksanakan mulai bulan Maret 2005 sampai dengan  Juli 2005.

### Metode Pengumpulan Data

Metode pengumpulan data yang dipakai dalam penyusunan dan penulisan penelitian ini adalah: 1) Data Primer, dan 2) Data Sekunder. Data Primer merupakan data yang dikumpulkan secara langsung dari objek yang diteliti. Adapun cara-cara yang dipakai untuk mengumpulkan data tersebut adalah: a)

Observasi, merupakan teknik pengumpulan data dengan cara mengadakan pengamatan langsung keobjek penelitian dengan mencatat secara sistematis data yang diperlukan dan melakukan pengamatan secara langsung yaitu kebagian perlengkapan pada Universitas Bina Darma Palembang, b) *Interview*, merupakan teknik pengumpulan data dengan cara mengadakan tanya jawab atau wawancara secara langsung kepada bagian perlengkapan guna mendapatkan data langsung dan informasi yang diperlukan. Sedangkan Data Sekunder merupakan mengumpulkan data dengan cara membaca buku literatur, majalah dan karangan ilmiah yang ada hubungan dengan penelitian ini.

### Metode Pengembangan Sistem

Metode pengembangan sistem yang dipakai adalah metode sekuensial linier untuk rekayasa perangkat lunak, yang sering disebut juga dengan "siklus hidup klasik" atau model air terjun (*water fall*), jenis pengembangan sistem ini merupakan dasar dari metode lain seperti model prototipe dan model RAD.

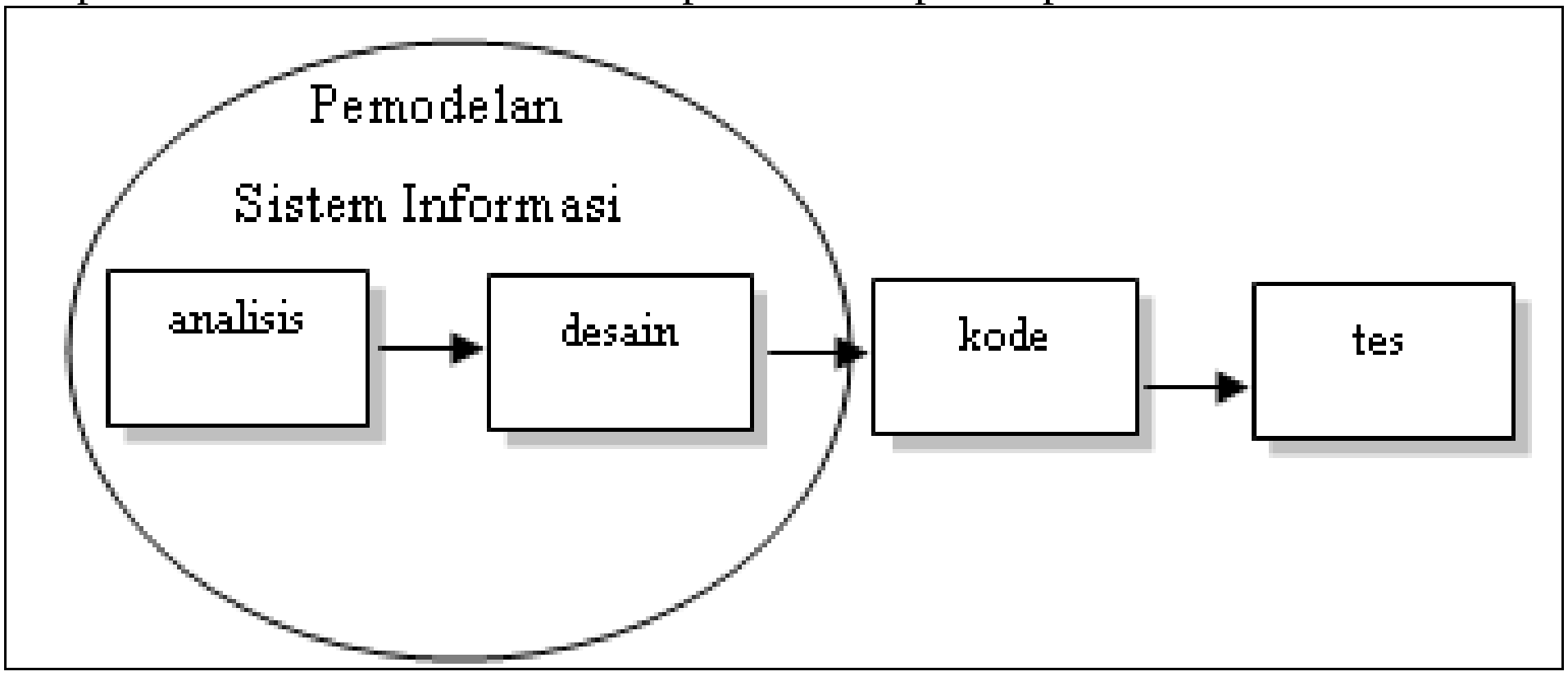

**Gambar 5. Model Sekuensial Linier**

Menurut Pressman (2002:36) model sekuensial linier melingkupi aktivitas sebagai berikut: 1) Analisis kebutuhan perangkat lunak, proses pengumpulan kebutuhan diintensifkan dan difokuskan, khususnya pada perangkat lunak. Untuk memahami sifat program yang dibangun, perekayasa perangkat lunak, 2) Desain, Desain perangkat lunak sebenarnya adalah proses multi langkah yang berfokus pada empat atribut yaitu: Struktur data, arsitektur perangkat lunak, representasi interface dan algoritma, 3) Kode, Pengkodean adalah penterjemahan desain kedalam bentuk bahasa mesin., dan 4) Tes (Pengujian), Sekali kode dibuat, pengujian program dimulai. Proses pengujian berfokus pada logika internal perangkat lunak, memastikan bahwa semua pernyataan yang diuji, dan pada ekternal fungsional yaitu mengarahkan pengujina untuk menemukan kesalahan-kesalahan dan memastikan bahwa input yang dibatasi akan memberikan hasil aktual yang sesuai dengan hasil yang dibutuhkan.

### Alat dan Bahan

Alat dan bahan yang digunakan pada penelitian ini terdiri atas: 1) perangkat keras (*hardware*), berupa: a) Personal komputer dengan *processor Intel Pentium III* 800 MHz, b) Monitor Samsung *Sync Master* 591s, c) Printer *Cannon iP1000*, d) *Mouse*, dan e) *Keyboard*, 2) Perangkat Lunak (*Software*), berupa: a) *Microsoft Windows XP* sebagai sistem operasi, b) *Microsoft Office XP*, dan c) *Microsoft Visual Studio 6.0*.

### Arsitektur Perangkat Lunak

Gambaran perangkat bantu membaca huruf arab malayu yang berupa beberapa proses di dalam pengerjaan atau alur jalannya simulasi itu sendiri.

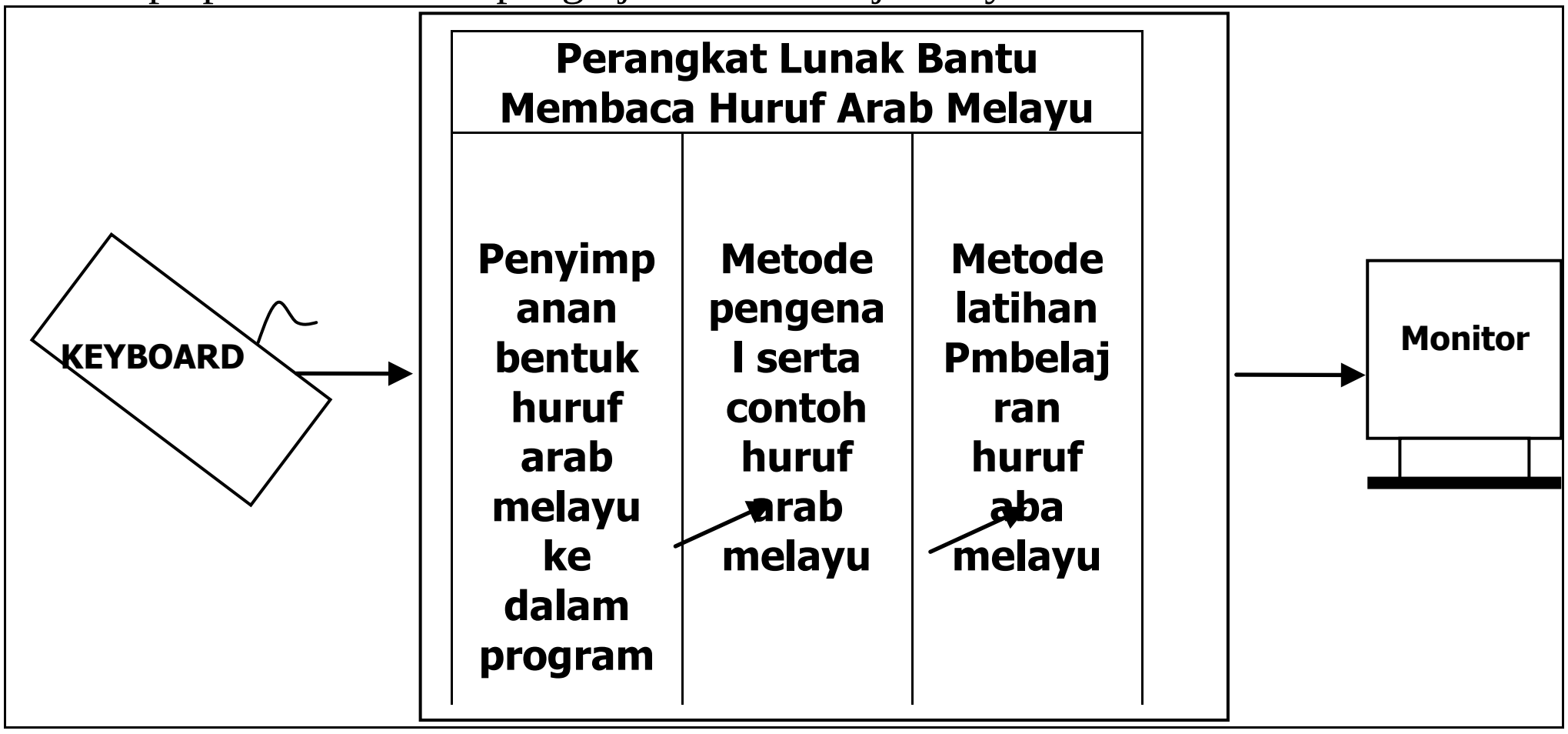

**Gambar 6. Arsitektur Perangkat Lunak Bantu Bacaan Huruf Arab Melayu**

**Rancangan Proses**

## 3.1 Spesifikasi Kemampuan Perangkat Lunak

Kemampuan perangkat lunak ini adalah dapat membantu dan mempermudah pemakai tentang cara penulisan huruf arab melayu hingga cara membaca dalam abjad latin. Adapun langkah-langkah cara penulisan huruf arab melayu, yaitu: 1) Didalam penulisan huruf arab melayu berbeda dengan penulisan huruf latin yang ditulis dari kiri ke kanan, penulisan huruf huruf arab melayu yaitu penulisan mulai kanan hingga kekiri, 2) Penulisan huruf arab melayu yang ditulis dalam keadaan berdiri sendiri, maka tidak boleh disambungkan dengan huruf-huruf lain selepasnya kecuali huruf arab melayu yang dibuat bersambung dengan huruf yang lainnya, dan 3) Berbeda dengan huruf-huruf arab melayu yang lainnya, huruf dal (د), dzal (ذ), ra (ر), zai (ز), waw (و) tidak boleh bersambung dengan huruf-huruf lain selepasnya, maka perlu ditulis seperti diawal sesuatu perkataan dan dirapatkan dengan huruf sebelumnya, contohnya: بوات

Setelah dapat mengetahui cara penulisan huruf arab melayu, maka selanjutnya mengetahui cara membaca huruf arab melayu yang dibaca dengan menggunakan ejaan dalam abjad latin, adapun cara pembacaannya berdasarkan kata atau kalimat yang dimasukan yang dibaca dimulai dari kanan hingga kekiri seperti pembacaan abjad latin ABCDEFGHIJKLMNOPQRSTUVWXYZ yang dimulai dari kiri hingga kekanan, contoh: buta (بوات).

**Proses Perangkat Lunak Membaca Huruf Arab Melayu**

Pada saat menjalankan perangkat lunak bantu membaca huruf arab melayu, pemakai akan disajikan tampilan dari perangkat lunak, berupa tampilan huruf-huruf arab melayu yang berdiri sendiri, berada diawal, tengah dan akhir kalimat. Adapun penggunaan perangkat lunak ini yaitu: 1) Pilihlah huruf-huruf arab melayu yang dimasukkan berdasarkan kalimat yang ingin dibuat, selanjutnya akan tampil huruf-huruf arab melayu berdasarkan pilihan tersebut, dan 2) Tekanlah huruf arab melayu tersebut dan kemudian akan tampil abjad latin berdasarkan pilihan huruf arab melayu, pilihlah abjad latin tersebut berdasarkan kalimat yang akan dibuat. Pada saat selesai menekan huruf arab melayu tersebut proseslah huruf arab melayu dan abjad latin tersebut menjadi untaian kata atau kalimat yang diinginkan.

**Proses Perangkat Lunak Materi Pembelajaran Huruf Arab Melayu**

Pada saat menjalankan materi dari penjelasan membaca huruf arab malayu, pemakai akan diberikan materi penjelasan huruf arab melayu mulai dari huruf-huruf yang digunakan pada perangkat lunak membaca huruf arab melayu, penjelasan huruf konsonan dan huruf vokal dan juga contoh dari penulisan huruf arab melayu berdararkan huruf yang ada pada huruf arab melayu.

**Langkah-Langkah dalam Pembuatan Perangkat Lunak Bantu**

Untuk mengetahui proses dalam pembuatan perangkat lunak bantu membaca huruf arab melayu, diperlukan langkah-langkah untuk proses pembuatannya: 1) Mencari fungsi yang terdapat dalam file sistem berbasis windows yang berjalan, kemudian mengaktifkan pemrograman *Microsoft Visual Basic* untuk pembuatan perangkat lunak bantu ini dan program perangkat lunak bantu membaca huruf arab melayu siap untuk dikerjakan, dan 2) Memasukkan huruf-huruf arab melayu satu-persatu serta pengertian huruf arab melayu, dengan ketentuan: a) Proses yang pembacaan huruf arab melayu dalam dikenal dalam sistem windows, serta b) Materi pembelajaran mulai pengenalan huruf arab melayu, huruf konsonan maupun vokal serta beberapa contoh dari cara penulisan huruf arab melayu.

## 4. PEMBAHASAN

**Struktur Perangkat Lunak**

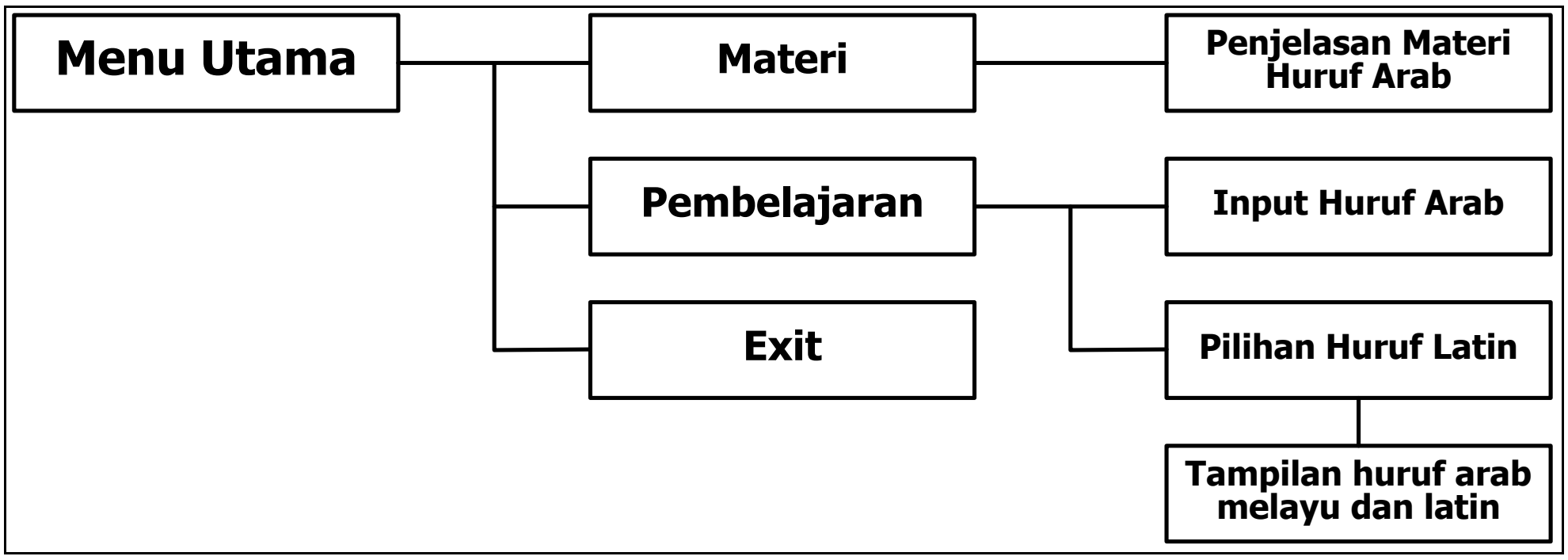

**Gambar 7. Struktur Perangkat Lunak**

Didalam rancangan struktur perangkat lunak ini untuk memulai perangkat lunak bantu membaca huruf arab melayu terdapat menu utama sebagai tampilan awal dari perangkat lunak bantu membaca huruf arab melayu, materi, pembelajaran, exit merupakan pilihan lanjutan untuk pada menu berikutnya.

Didalam pilihan Materi berfungsi sebagai penjelasan huruf arab melayu. *Form* pembelajaran untuk memulainya masukkan *input* huruf arab melayu disertai dengan pilihan huruf latin yang keluarannya berupa tampilan kata / kalimat huruf arab melayu disertai dengan huruf latin.

## Form Menu Utama

Form ini menjelaskan cara pemakaian perangkat lunak bantu membaca huruf arab melayu. Dengan menu utama yang mempnnyai latar belakang serta 3 (tiga) buah *command* yang digunakan untuk mengeksekusi program, pembelajaran huruf arab melayu, materi huruf arab melayu, dan keluar dari program (Gambar 6).

Selanjutnya pada rancangan perangkat lunak yang pembelajaran huruf arab melayu, terdapat tampilan dari huruf arab melayu, berdasarkan pilihan huruf yang berdiri sendiri, huruf awal, huruf tengah dan huruf akhir, serta pada masing-masing huruf arab melayu tersebut terdapat *label* yang menerangkan bahwa huruf tersebut bunyi dasarnya seperti tampilan huruf yang ada bawah label tersebut.

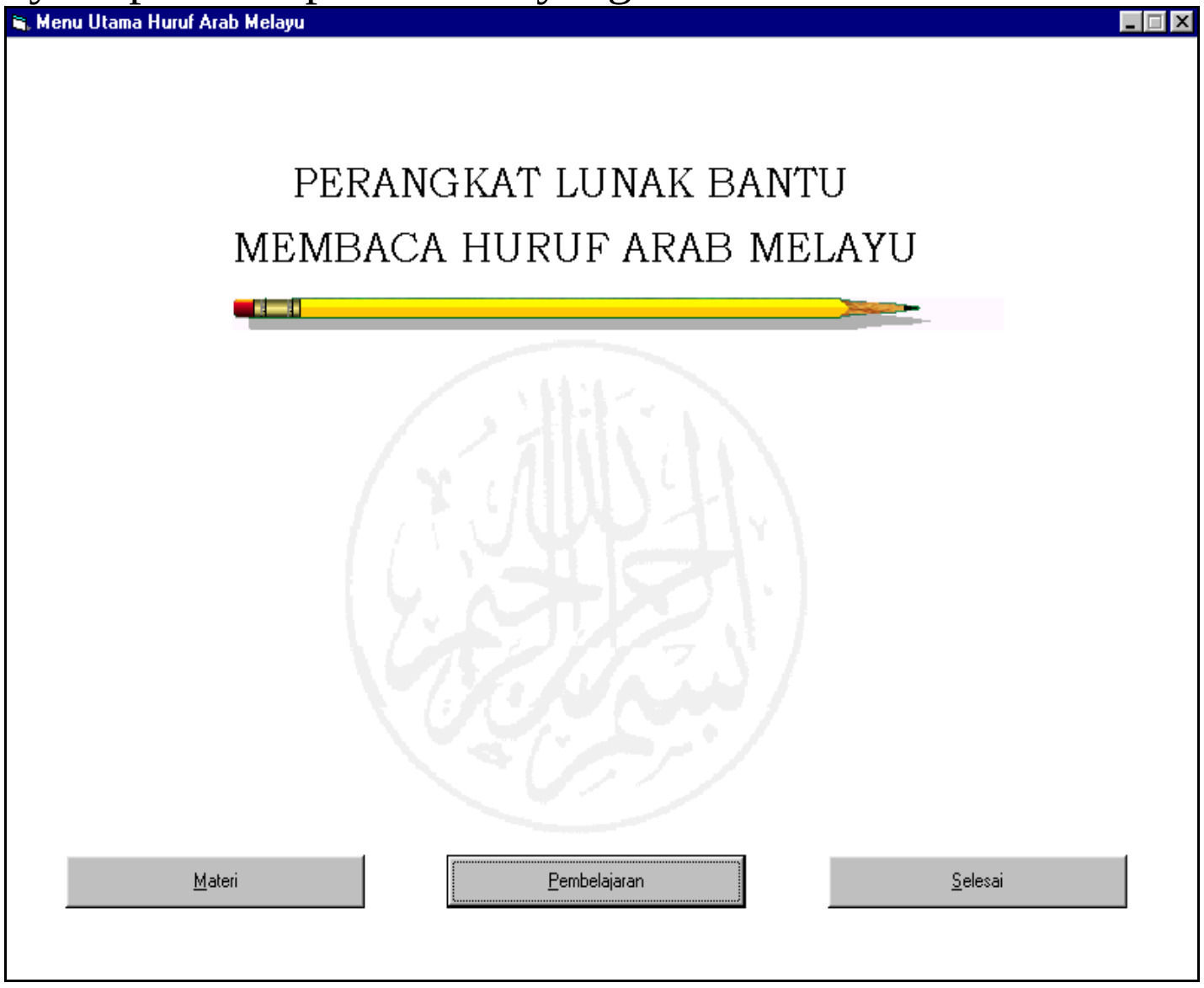

**Gambar 8. Menu Utama**

## Form Materi

Untuk melihat isi form materi, klik tombol materi, maka akan nampak form seperti gambar 9. form ini berisi penjelasan teknik penulisan huruf arab melayu, huruf vokal, dan konsonan.

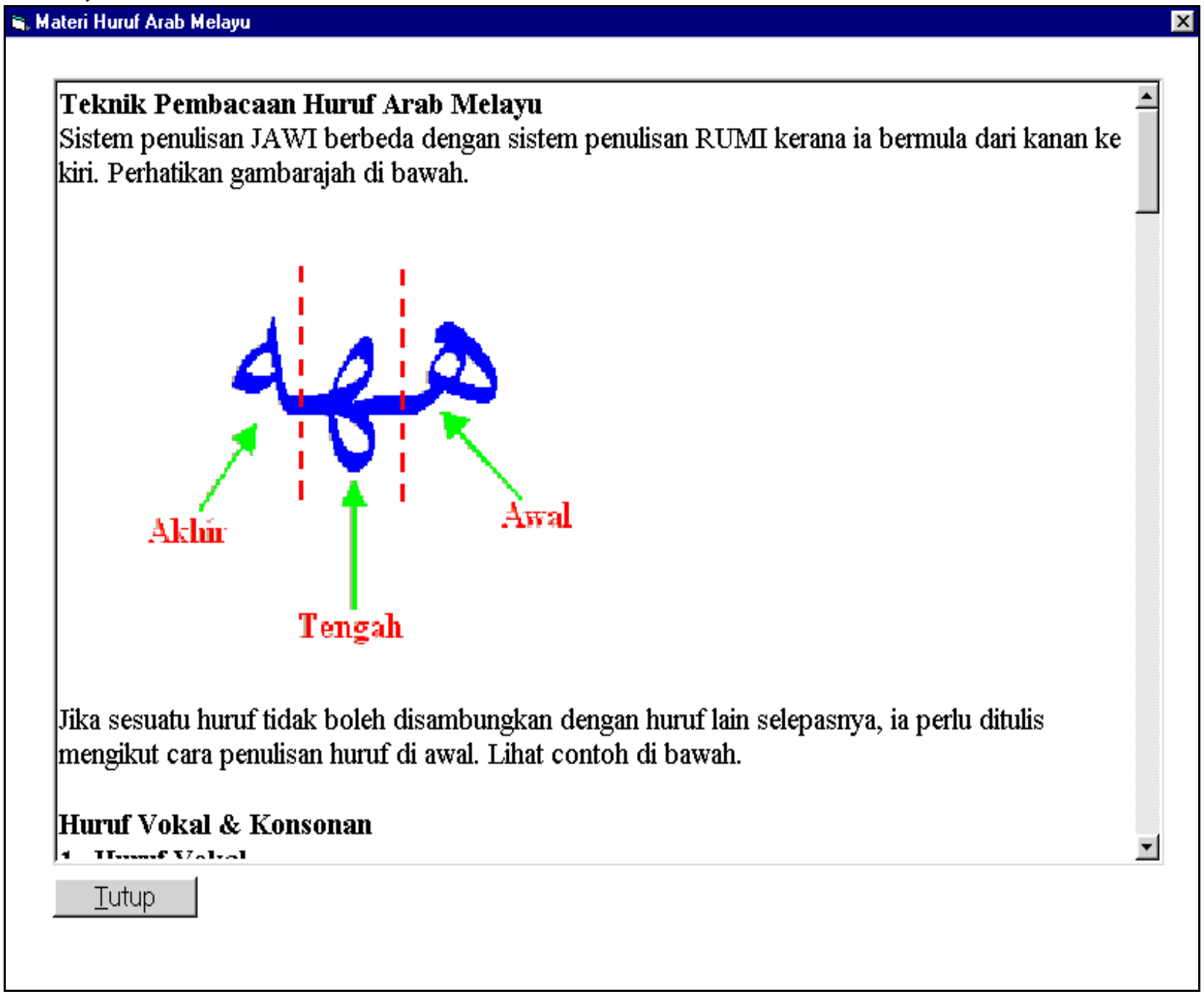

**Gambar 9. Form Materi**

## Form Pembelajaran

Selanjutnya pada rancangan perangkat lunak yang pembelajaran huruf arab melayu, terdapat tampilan dari huruf arab melayu, berdasarkan pilihan huruf yang berdiri sendiri, huruf awal, huruf tengah dan huruf akhir, serta pada masing-masing huruf arab melayu tersebut terdapat *label* yang menerangkan bahwa huruf tersebut bunyi dasarnya seperti tampilan huruf yang ada bawah label tersebut.

Pada saat kita ingin memasukkan kata abjad dari huruf arab melayu tersebut, disini terdapat 2 buah *text* yang berfungsi sebagai tampilan dari kata atau kalimat huruf arab melayu yang pada penulisannya dimulai dari sebelah kanan, dan juga *text* untuk abjad latin yang penulisannya dimulai dari sebelah kiri.

Selain terdapat tampilan *text* juga terdapat 8 buah *command* yang masing-msin berfungsi yaitu *command* proses untuk mengeksekusi abjad huruf arab melayu yang diletakan pada text huruf arab melayu dan abjad latin, *command* berdisi sendiri sebagai penunjuk dari huruf arab melayu yang berdisi sendiri, *command* awal sebagai penunjuk dari huruf arab melayu yang merupakan penulisan huruf awal dari huruf arab melayu, *command* tengah merupakan penunjuk bahwa abjad ini adalah huruf tengah diantara 2 buah huruf arab melayu, *command* akhir merupakan penunjuk akhir dari sebuah kata dari huruf arab melayu, *command* pembelajaran yang digunakan sebagai penjelasan dari huruf arab melayu disertai dengan contoh penulisan dan pembacaan dari huruf arab melayu, *command* menu utama digunakan untuk kembali pada menu utama dan *command* selesai digunakan untuk keluar dari program ini.

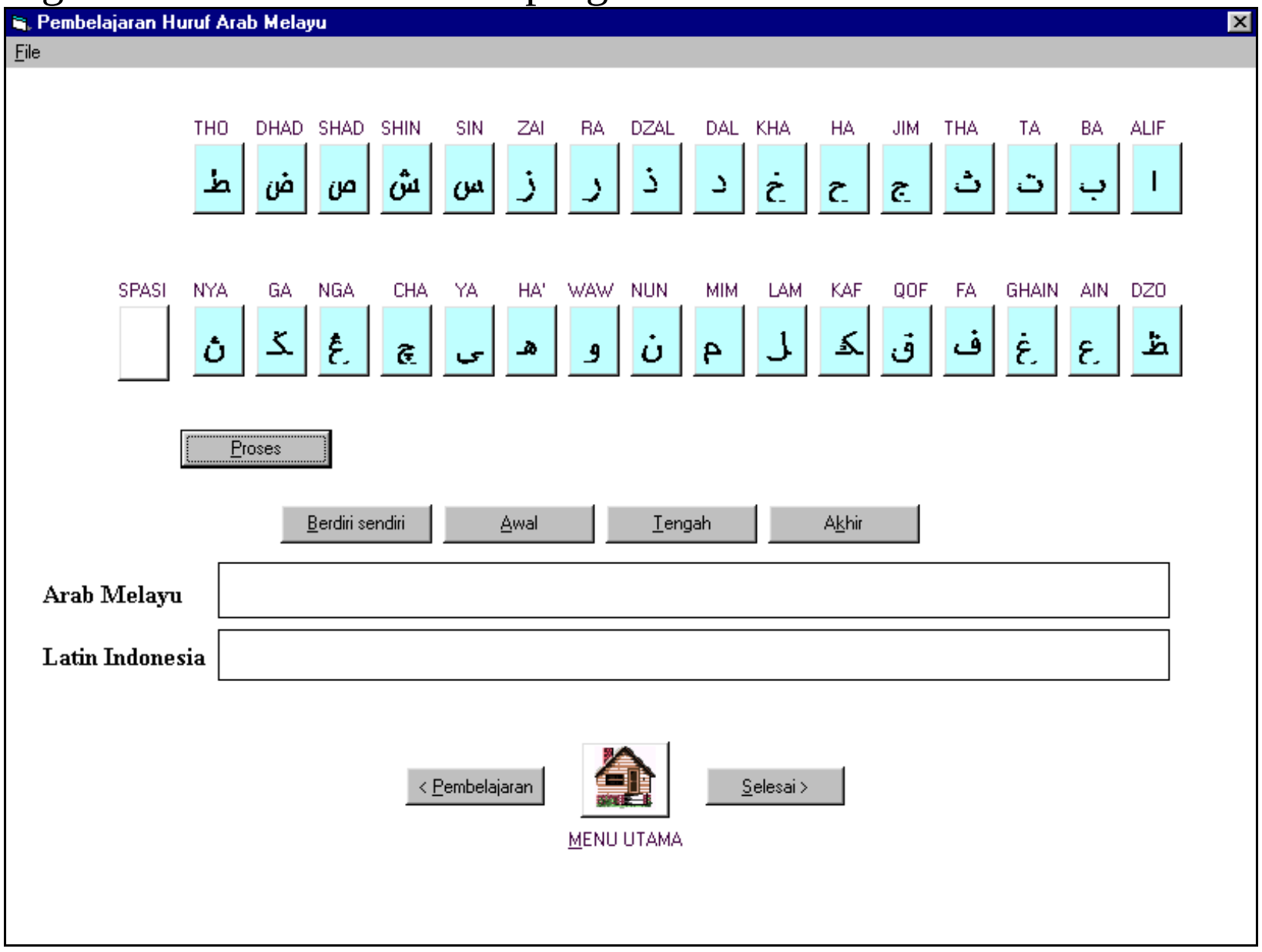

**Gambar 10. Pembelajaran Huruf Arab Melayu**

Untuk menggunakan perangkat lunak ini sebelumnya pastikan terlebih dahulu telah mempelajari materi dari huruf arab melayu, yang maksudnya supaya dapat menggunakan tampilan pembelajaran huruf arab melayu dengan baik.

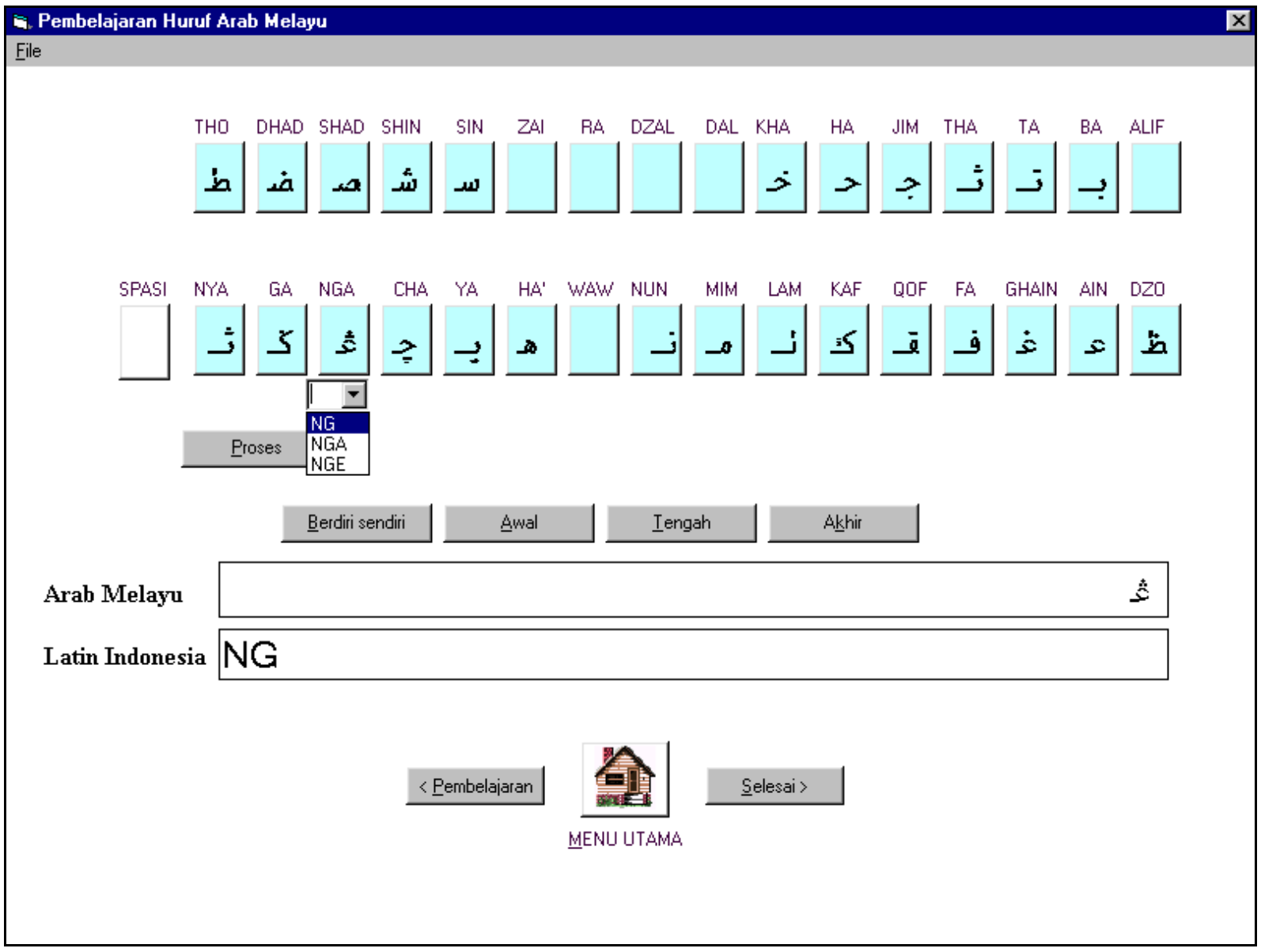

**Gambar 11. Tampilan Pembelajaran Berdasarkan Pilihan Huruf Awal Huruf Arab Melayu**

Dalam membuat kata atau kalimat huruf arab melayu ini tekanlah pilihan *command* berdiri sendiri, *command* awal, *command* tengah dan *command* akhir untuk membentuk abjad dari huruf arab melayu ini. Pada saat penekanan huruf Nga (y) awal maka akan tampil pilihan huruf latin yang akan ikut serta tampil bersamaan dengan huruf arab melayu tersebut, tekanlah huruf latin pilihan tersebut, kemudian tekanlah command proses untuk menampilkan kedalam huruf arab melayu dan latin Indonesia.

Untuk penekanan huruf selanjutnya, tekanlah huruf alif akhir (#) dan pilihlah huruf latin (A) yang digunakan sesuai dengan pilihan huruf arab melayu yang digunakan sebagai penunjuk dari huruf alif akhir yang dipilih tesebut yang terdapat pada gambar 12.

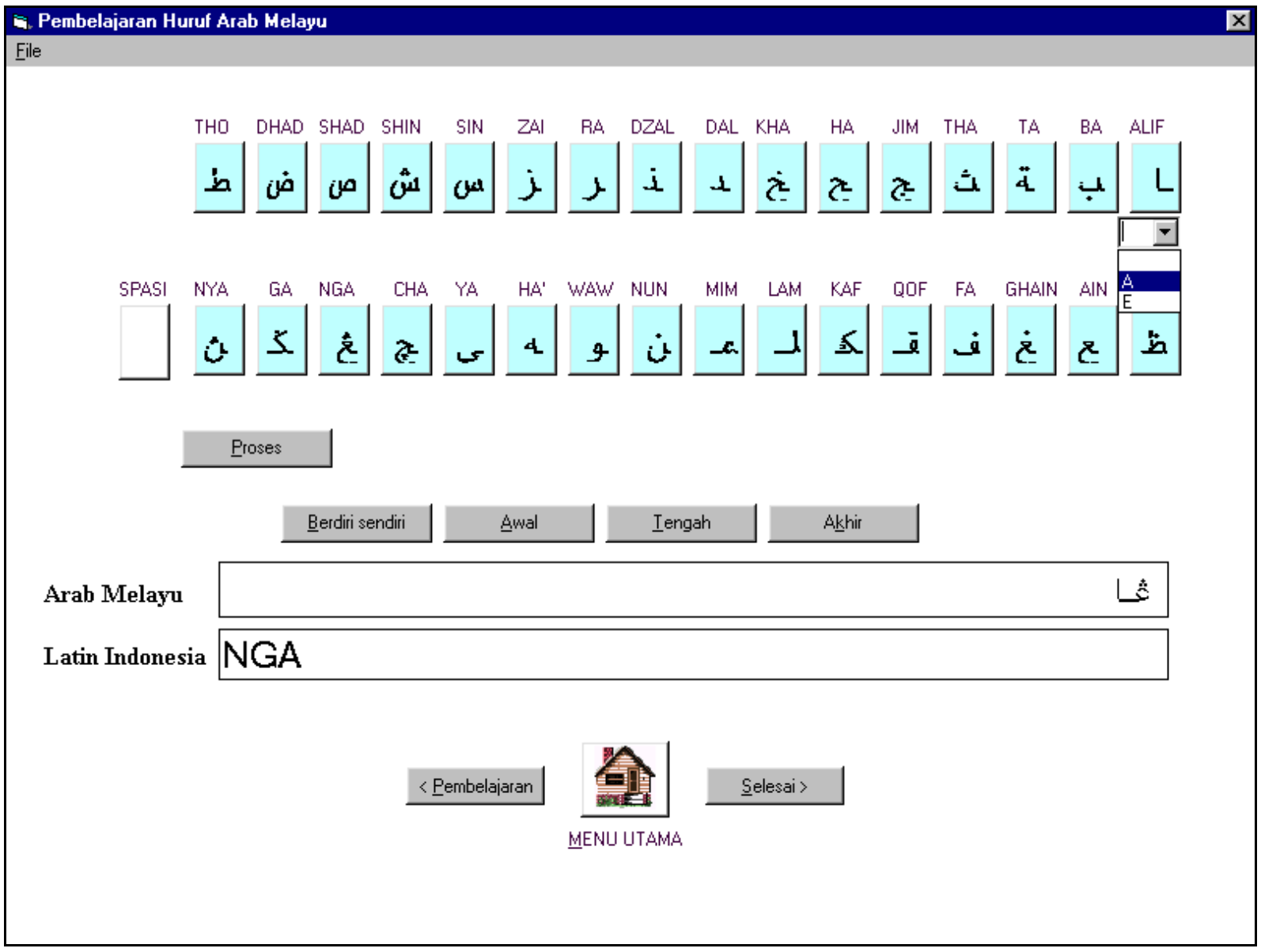

**Gambar 12. Tampilan Pembelajaran Berdasarkan Pilihan Huruf Akhir Huruf Arab Melayu**

Dalam membuat kata atau kalimat huruf arab melayu ini tekanlah pilihan *command* berdiri sendiri, *command* awal, *command* tengah dan *command* akhir untuk membentuk abjad dari huruf arab melayu ini. Pada saat penekanan huruf Nga (y) awal maka akan tampil pilihan huruf latin yang akan ikut serta tampil bersamaan dengan huruf arab melayu tersebut, tekanlah huruf latin pilihan tersebut, kemudian tekanlah command proses untuk menampilkan kedalam huruf arab melayu dan latin indonesia, untuk memasukkan huruf yang lainya lakukan seperi proses penekanan huruf Nga (y) awal tersebut. Ulangi langkah tersebut untuk menjadi kalimat yang diinginkan tersebut.. Untuk memulai dari awal pembuatan huruf arab melayu ini tekanlah Ctrl+N. untuk keluar dari proses yang sedang berlangsung tekanlah atau kembali ke menu utama tekanlah *command* selesai atau *command* menu utama.

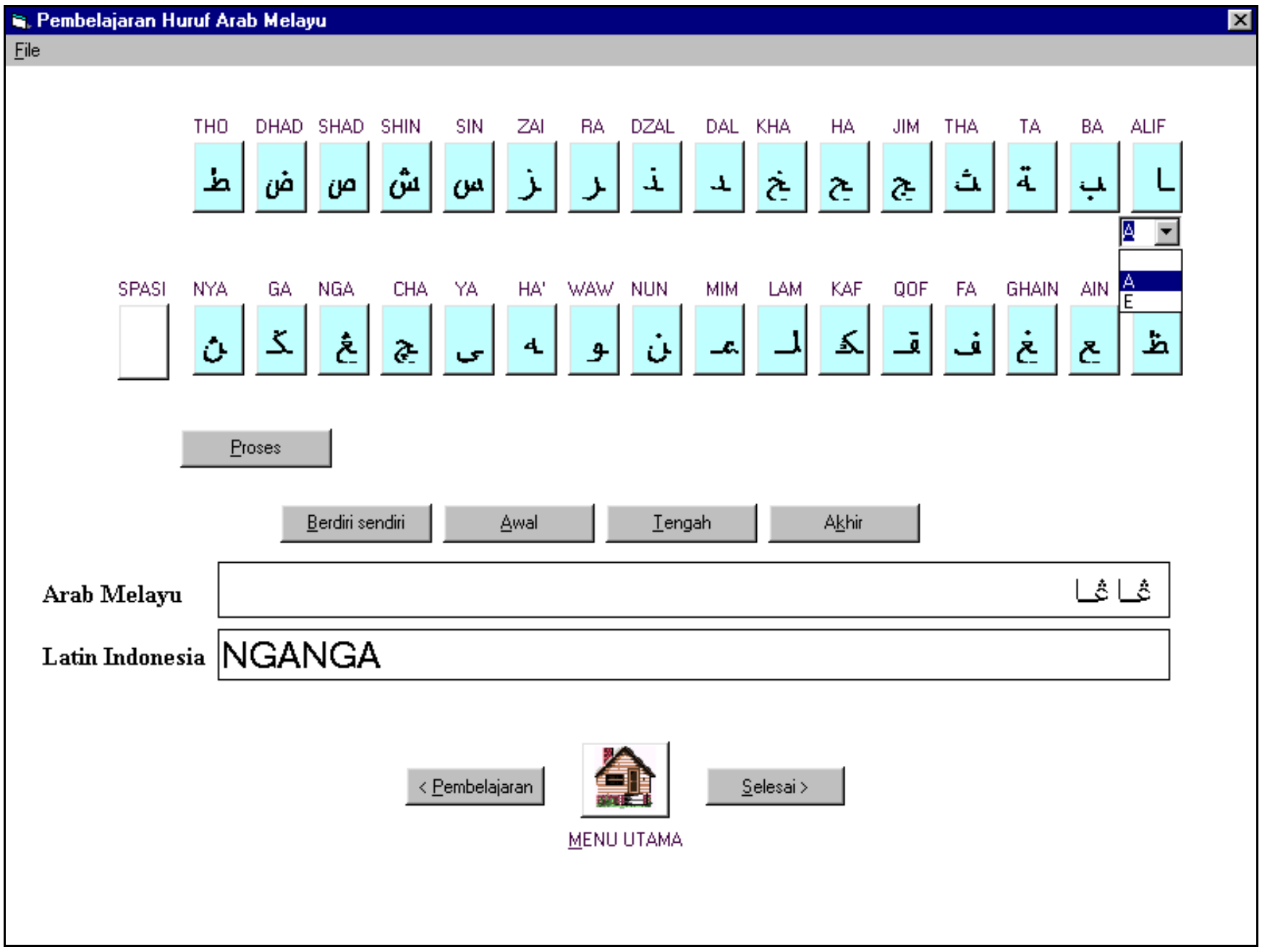

**Gambar 13. Tampilan Pembelajaran Huruf Arab Melayu**

## 5. KESIMPULAN

Berdasarkan hasil penelitian ini dapat diambil beberapa kesimpulan tentang perangkat lunak Bantu mengenal huruf arab melayu ke bentuk huruf latin indonesia sebagai berikut:

1) Perangkat lunak bantu ini sangat mudah digunakan, karena Anda tidak perlu menuliskan huruf Arab secara langsung, namun terlah disediakan sejumlah pilihan dari daftar huruf-huruf-nya.
2) Hasil dari perangkat lunak bantu membaca huruf awab melayu ini digunakan untuk mempelajari cara penulisan dan cara membaca huruf arab melayu dan kemudian diubah menjadi huruf latin.
3) Perangkat lunak bantu membaca huruf arab melayu ini akan menjadi kata atau kalimat berdasarkan pilihan huruf arab melayu dan latin.

## DAFTAR RUJUKAN

__________, 2005, *Huruf Jawi*, (OnLine), (http://id.wikipedia.org/wiki/Huruf_Jawi, diakses pada Desember 2005), Wikipedia Foundation Inc., USA.

__________, 2005, *Tulisan Jawi*, (OnLine), http://ms.wikipedia.orgwiki/Tulisan_Jawi/TulisanJawi-Wikipedia.htm, diakses pada Desember 2005), Wikipedia Foundation Inc., USA.

Abdillah, Leon, Andretti, 2006, *Pemrograman II*, Universitas Bina Darma Press, Palembang.

Hartono, Jogiyanto, 1992, *Pengenalan Komputer*, Andi Offset, Yogyakarta.

Pressman, Roger S., 2002, *Rekayasa Perangkat Lunak: Pendekatan Praktisi*, Andi, Yogyakarta.

Suryo, Kusuma, Aryo, 2000, *Buku Latihan Microsoft Visual Basic 6.0*, Elex Media Komputindo, Jakarta.

Tim Penyusun, 1997, *Kamus Besar Bahasa Indonesia*, Balai Pustaka, Jakarta.